\documentclass[reprint, superscriptaddress, showpacs, amsmath,amssymb,aps,prl]{revtex4-1}

\usepackage{graphicx}
\usepackage{subfigure}
\usepackage{dcolumn}
\usepackage{bm}
\usepackage{longtable}
\usepackage{amssymb}

\begin{document}
\title{Transition Path, Quasi-potential Energy Landscape and Stability of Genetic Switches}

\author{Cheng Lv}
\affiliation{School of Physics, Peking University, Beijing 100871, China}
\author{Xiaoguang Li}
\affiliation{LMAM and School of Mathematical Sciences, Peking University, Beijing 100871, China}
\author{Fangting Li}
\email{li\_fangting@pku.edu.cn}
\affiliation{School of Physics, Peking University, Beijing 100871, China}
\affiliation{Center of Quantitative Biology, Peking University, Beijing 100871, China}
\author{Tiejun Li}
\email{tieli@pku.edu.cn}
\affiliation{LMAM and School of Mathematical Sciences, Peking University, Beijing 100871, China}
\affiliation{Beijing International Center for Mathematical Research, Beijing 100871, China}

\begin{abstract}
One of the fundamental cellular  processes governed by genetic regulatory networks in cells is the transition among different states under the intrinsic and extrinsic noise. Based on a two-state genetic switching model with positive feedback, we develop a framework to understand the metastability in gene expressions. This framework is comprised of identifying the transition path, reconstructing the global quasi-potential energy landscape, analyzing the uphill and downhill transition paths, etc. It is successfully utilized to investigate the stability of genetic switching models and fluctuation properties in different regimes of gene expression with positive feedback.
%Our analytical results agree well with Monte Carlo simulations.
%We also show that the gradient-curl type decomposition of the deterministic drift term may hold only near the %stationary fixed points.
The quasi-potential energy landscape, which is the rationalized version of Waddington potential,  provides a quantitative tool to understand the metastability in more general biological processes with intrinsic noise.
\end{abstract}


\maketitle

%\section{}
In a cell, the reactions underlying gene expression involve small numbers of molecules, such as DNA, mRNAs, and transcription factors, so the stochasticity in gene regulation process is inevitable even  under constant environmental conditions \cite{Munsky2012,Balazsi2011}. Recent progresses in single-cell observations and analysis demand new quantitative methods to characterize the regulatory mechanism in both prokaryotes \cite{Elowitz2002} and eukaryotes \cite{OShea2004} from the cellular stochasticity.

 Previous kinetic studies of cellular stochasticity have been formulated by using the generating function \cite{GFPapers}, system size expansion \cite{Paulsson2004,Kampen1981}, large deviation theory \cite{Wang2010,LDTPapers}, or by employing WKB approximation to the chemical master equations (CMEs) \cite{Walczak2005,Wolynes2005,Assaf2011}, etc. However, only few of them take account of transcriptional noise explicitly. Some recent studies have shown that correlations between mRNA and protein levels do not always perform equally well in revealing genetic regulatory relationships \cite{Gandhi2011,Taniguchi2010},  and the consideration of mRNA has intensive effect on the switching times \cite{Mehta2008,Golding2010}. Since Waddington's ``epigenetic landscape" proposed in 1957 \cite{Waddington1957}, the energy landscape have been widely used to provide pictorial illustration of the dynamics and evolution of genetic regulatory systems \cite{Zhang2006,Wang2010,Balazsi2011}. Thus it is important and natural to develop a general methodology which can effectively determine the key features of a gene expression system, such as constructing the corresponding ``Waddington potential'', identifying the transition paths between metastable states and computing the transition rates, etc.

In this letter, we present a framework to understand the metastability of the genetic switches in gene expression based on the large deviation theory for Markov processes \cite{Weiss,Freidlin1998,Varadhan}. By explicitly taking into account the mRNA noise, we obtain the most probable transition paths for {\it off}-to-{\it on} and {\it on}-to-{\it off} genetic switches through the geometric minimum action method (gMAM) \cite{Heymann2008}. Furthermore, we reconstruct the global quasi-potential energy landscape, which is the rationalized version of the Waddington potential,  via the computation of the local quasi-potentials. We analyze the properties of transition paths,  discuss their relation to the quasi-potential energy landscape and the deterministic dynamics, and compare the differences between our approach and others in previous literatures. Our analytical results agree well with the Monte Carlo simulations. We also discuss the choice of the system size and its finite effect via the transition path theory (TPT) \cite{TPT}.  From the authors' opinion, this framework is generally applicable for studying transitions between stable-saddle-stable fixed points with jump type noise generated by Gillespie's birth-death dynamics \cite{Gillespie1977}.  It is successfully utilized to investigate the stability of genetic switching models and fluctuation properties in different regimes of gene expression with positive feedback, which leads to interesting biological insights.
\begin{figure}[htbp]
\includegraphics{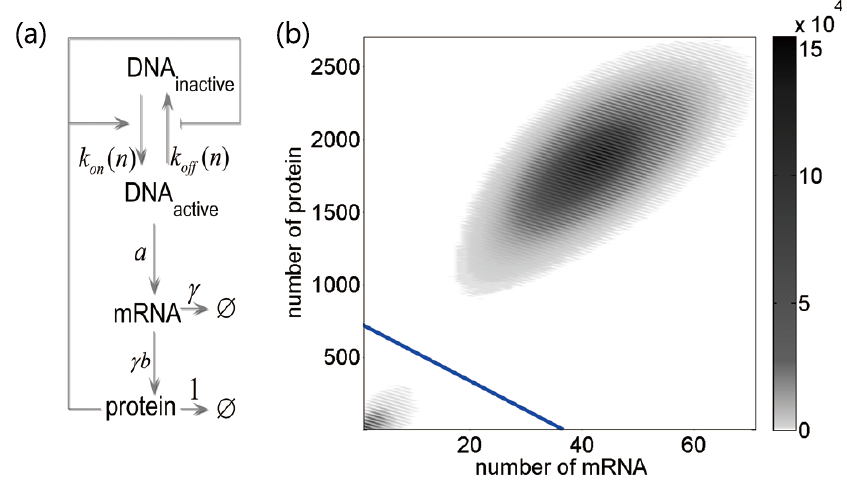}
%\scalebox{0.27}{\includegraphics{Gillespie_1.jpg}}
\caption{\label{model}(color online). (a) Model for gene-expression switching network following Assaf et al \cite{Assaf2011}. Promoter transitions are regulated by the feedback functions \(k_{on}\) and \(k_{off}\). Transcription, translation and decay of mRNA and protein are modeled as first-order reactions with rates \(a, \gamma b, \gamma\) and \(1\) (rates are rescaled by the protein decay rate). (b) Visiting frequency distribution of mRNA and protein with a long time Monte Carlo simulation (switches occur 2000 times). Darkness of points  shows the number of visits with suitable smoothing. The blue solid line is the separatrix between two basins of attraction of on and off states.  In (b), \(K=ab=2400, b=22.5, h=2, n_{50}=1000, k_0^{\min}=k_1^{\min}=a/100, k_0^{\max}=k_1^{\max}=a\) and \(\gamma=2.\)}
\end{figure}

Let us focus on a two-state gene expression model in Fig. \ref{model}(a) following the previous study \cite{Assaf2011}, where the transitions of active and inactive promoter states are controlled by the protein via a positive feedback. The transition rates between active and inactive states are \begin{math}k_{on}(n)\equiv f(n)\end{math} and \begin{math}k_{off}(n)\equiv g(n)\end{math}, where \(n\) is the protein copy number, {\it on} and {\it off} denote the high and low states of protein and mRNA respectively. Here the transition rates are in the form of Hill-function as
\begin{math}
f(n)=k_0^{\min}+(k_0^{\max}-k_0^{\min})n^{h_1}/(n_{50}^{h_1}+n^{h_1})
\end{math} and \begin{math}
g(n)=k_1^{\max}-(k_1^{\max}-k_1^{\min})n^{h_2}/(n_{50}^{h_2}+n^{h_2})
\end{math}, where \(n_{50}\) is the curve's midpoint, and \(h_1=h_2=h\) are chosen for simplicity.

A deterministic description of the genetic switching model in Fig. \ref{model} is given by the equations
\begin{equation}\label{ODE_1}
\dot{M}=af(N)/[f(N)+g(N)]-\gamma M, \ \dot{N}=\gamma bM-N,
\end{equation}
through quasi-steady state approximation, where \(M,N\) denote the mean number of mRNAs and proteins, respectively.
%The term \(f(N)/[f(N)+g(N)]\)
%originates from the quasi-stationary approximation which represents the probability of the promoter in its %active state.
The corresponding stochastic description is governed by the CMEs \cite{Kampen1981}
\begin{align}\label{CMEs}
&\dot P_{m,n}=g(n) Q_{m,n}-f(n) P_{m,n}+\mathbf A P_{m,n},\\
&\dot Q_{m,n}=-g(n) Q_{m,n}+f(n) P_{m,n}+[\mathbf A+a(\mathbf E_m^{-1}-1)] Q_{m,n}.\nonumber
\end{align}
where \( P_{m,n}\) denotes the probability distribution function (PDF) of inactive DNA/promoter state with $m$ mRNAs and $n$ proteins at time t, while \( Q_{m,n}\) denotes the active DNA/promoter state.  Here we employ the notation for raising operator \(\mathbf E_n^j\) acting on $f(n)$ as \(\mathbf E_n^jf(n)=f(n+j)\), and \(\mathbf A\equiv(\mathbf E_n^1-1)n+\gamma(\mathbf E_m^1-1)m+\gamma bm(\mathbf E_n^{-1}-1)\) is a birth-death operator related to the inactive promoter.  Equations (\ref{ODE_1}) exhibit the bistability but ignores the noise. Under the deterministic description, once the system settles in one of its two attractive fixed points, it will stay there forever. However, in the presence of intrinsic noise, the system will fluctuate around its attractive fixed points and switch between its two metastable states on a large timescale (see Fig. \ref{model}(b)). In the regime of interest (see also \cite{Assaf2011}), we fix
parameters $\gamma,b$ and let $a=Kb^{-1}$. Here $K$ plays the role of system size \cite{Weiss,Kampen1981}. We will let $K$ goes to infinity and fix the ratios $n_{50}/K, k_{0}^{\min}/K,k_{0}^{\max}/K,k_{1}^{\min}/K,k_{1}^{\max}/K$ in the limiting process. This choice gives the rationale that in which sense the ODEs (\ref{ODE_1}) is the mean field limit of the CMEs (\ref{CMEs}) (cf. the Supplementary Material ({\it SM})).

The large deviation theory gives a good quantitative description of rare events \cite{Weiss, MAM, Heymann2008}.  The most probable transition path is given by minimizing an action functional characterized  by a Lagrangian $L$. For the Gillespie's birth-death dynamics, $L$ has no closed form and only its dual Hamiltonian can be given as
\begin{equation}\label{jump H}
H(\boldsymbol{x,p})=\sum_{j=1}^Na_j(\boldsymbol x)(e^{\boldsymbol{p\cdot \nu}_j}-1),
\end{equation}
where \(a_j\) is the rate (or propensity) and \(\boldsymbol \nu_j\) is the state-change (or stoichiometric) vector, \(j=1,2,\ldots,N\). However, this description has difficulty when taking DNA into consideration, for normally there are too few DNA copies in a living cell that the straightforward application of the above
formulation is not valid.

%This is exactly the  Hamiltonian corresponds to the Hamilton-Jacobi equation satisfied by the quasi-potential to %be introduced through WKB asymptotics, which will be utilized in the continued text (cf. also \cite
%{Dykman1994,Dykman1995}).

%We are interested in the sufficiently long time behavior of the system and hence the stationary distribution
%\(\dot P_{m,n}=\dot Q_{m,n}=0\) in Eqs.(\ref{CMEs}). To take DNA and mRNA noise into account,
Now we adopt Assaf et al's approach \cite{Assaf2011} by considering the stationary distribution where \(\dot P_{m,n}=\dot Q_{m,n}=0\) and eliminating \(Q_{m,n}\) to obtain
\begin{equation}\label{CME}
0=\{\mathbf A+g(n)^{-1}[\mathbf A+a(\mathbf E_m^{-1}-1)][f(n)-\mathbf A]\}P_{m,n}.
\end{equation}
%
%The small parameter $\varepsilon$ in Eq.\eqref{F-W} corresponds to $K^{-1}$,
 Define the concentration variable \(x=m/K\), \(y=n/K\) and the quasi-potential \(S(x,y)\) \cite{Freidlin1998}, we plug WKB ansatz
\begin{equation}\label{WKB}
P_{m,n}\equiv P(x,y)\sim \exp[-KS(x,y)]
\end{equation}
into Eq. (\ref{CME}) to get a stationary Hamilton-Jacobi equation \(H(x,y,\partial_xS,\partial_yS)=0\) to the leading order with the Hamiltonian
\begin{equation}\label{Hamiltonian}
H=A+\tilde g(y)^{-1}[A+b^{-1}(e^{p_x}-1)][\tilde f(y)-A],
\end{equation}
where the function \(A=A(x,y,p_x,p_y)=y(e^{-p_y}-1)+\gamma x(e^{-p_x}-1)+\gamma bx(e^{p_y}-1)\), and the feedback functions are now rescaled as \(\tilde f(y)=f(y)/K, \tilde g(y)=g(y)/K\). The classical Hamilton-Jacobi theory enables one to solve the quasi-potential $S(x,y)$ with different formulations such as the variational methods, integrating Hamiltonian dynamics or directly solving PDEs, in which we have the connection \(p_x=\partial_xS\) and \(p_y=\partial_yS\) between the momenta and the quasi-potential \cite{Goldstein}. It is worth asking whether the choice of the large parameter $K$ affects the final results since any choice is artificial in practice.  An affirmative answer is given in {\it SM} that only the scaling matters and  the final systems are equivalent with respect to different choices of the large parameter $K$.

\begin{figure}[htbp]
\includegraphics{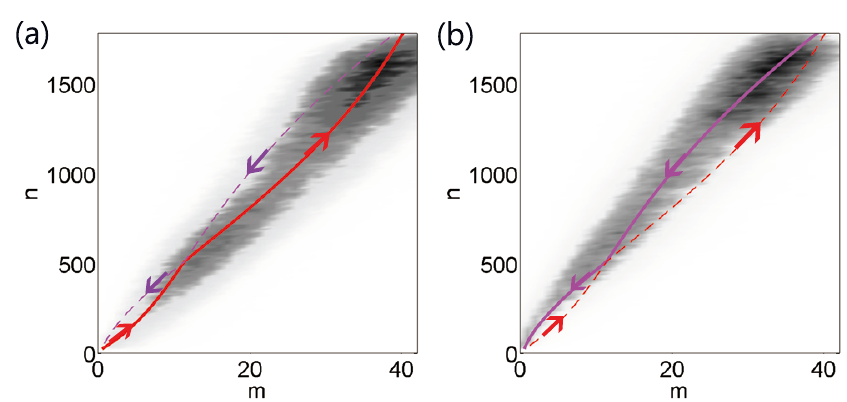}
\caption{\label{purepath}(color online). Switching paths (a) from {\it off} to {\it on} state (red solid curve) and (b) from {\it on} to {\it off} state (purple solid curve) and MC simulations for both switching trajectories. We take the  two stable fixed points in the deterministic dynamics as the starting and ending points. Darkness of the shading points represents the number of visits for reactive trajectories with smoothing. The results are obtained from 1000 independent long time MC simulations. Here, the parameters are the same as in Fig. \ref{model}.}
\end{figure}

Now we use the powerful gMAM algorithm \cite{Heymann2008} to compute the quasi-potential by minimizing the action functional  with the obtained Hamiltonian (\ref{Hamiltonian}) (some details about the gMAM can be referred to {\it SM}). We choose the two stable points solved by Eqs. (\ref{ODE_1}) as our starting and ending points, then compute the switching paths starting from either of the two states (see Fig. \ref{purepath}). These switching paths predict well with the switching trajectories obtained from the MC simulations.

Figure \ref{purepath} shows clearly that when switch occurs, the switching trajectory prefers to be around the most probable path characterized by the Hamiltonian (\ref{Hamiltonian}). Furthermore, the {\it off}-to-{\it on} and {\it on}-to-{\it off} paths are not identical, suggesting that the switching paths are irreversible, which is different from the original Waddington picture \cite{Waddington1957}. The irreversibility is fundamental
in chemical reaction kinetics due to the non-gradient nature of the dynamical system. Figure \ref{purepath} also tells that both switching paths pass through the same bottleneck, i.e. the saddle point given by Eqs. (\ref{ODE_1}). To further characterize the switching path, we note that the transition path is also given by the Hamilton's equations \(\dot {\boldsymbol x}=\nabla_{\boldsymbol p}H\), \(\dot {\boldsymbol p}=-\nabla_{\boldsymbol x}H\), where $\boldsymbol x=(x,y), \boldsymbol p=(p_{x},p_{y})$. Based on the fact $H(x,y,0,0)\equiv 0$, we obtain $\nabla_{\boldsymbol x}H\equiv \boldsymbol 0$ when $\boldsymbol p=\boldsymbol 0$. At the saddle point in any transition path, we have $\boldsymbol p= \boldsymbol 0$ \cite{Heymann2008}, and thus $\boldsymbol p\equiv \boldsymbol 0$ along the whole downhill path. With this result we obtain the downhill equations \(\dot {\boldsymbol x}=\nabla_{\boldsymbol p}H(\boldsymbol {x,0})\), which exactly corresponds to the deterministic dynamics \eqref{ODE_1} with a renormalization factor $\tilde{g}(y)/[\tilde{f}(y)+\tilde{g}(y)]$ of time since this only involves the DNA inactive state. This fact explains that after climbing the saddle point the biological system relaxes to its attracting state fast without cost any action.
The uphill path is then given by \(\dot {\boldsymbol x}=\nabla_{\boldsymbol p}H(\boldsymbol x,\nabla_{\boldsymbol x} S)\) which is in general different from the downhill path unless the system is of gradient type.

\begin{figure}[htbp]
\includegraphics{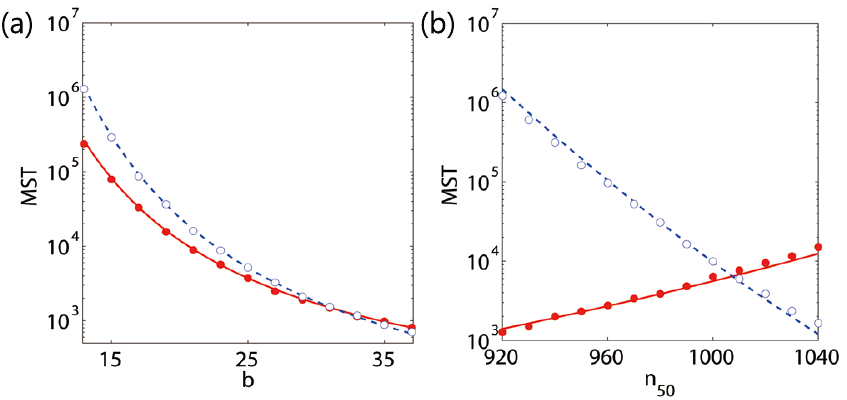}
\caption{\label{MST:f}(color online). The Mean Switching Time (MST) from both states as a function of (a) translation rate \(b\) (\(\gamma\) is hold constant) and (b) Hill-type function curve's midpoint \(n_{50}\).
The gMAM results with numerical prefactor from off to on states (red solid line) and from on to off states (blue dashed line), while MC simulations (\(\bullet\)) and (\(\circ\)), respectively. Prefactor from off to on state were (a) 31.08, (b) 27.61 and from on to off states (a) 4.82, (b) 7.94. Here \(h=2, K=ab=2400, k_0^{\min}=k_1^{\min}=a/100, k_0^{\max}=k_1^{\max}=a, \gamma=2.\) And for (a) \(n_{50}=1000,\) (b) \(b=22.5.\)}
\end{figure}

Besides giving the minimum action path, gMAM also contributes the action value at each point in the path, which enables us to calculate the mean switching time (MST) \(\tau\) from either stable state. Denote the quasi-potential energy barrier $\Delta S_{\rm on} = S_{0}-S_{\rm on}$, where $S_{0}$ is the action at the saddle point and $S_{\rm on}$ is the action at the {\it on} state. Then the MST $\tau_{\rm on}$ from {\it on}-to-{\it off} transition can be roughly estimated from an asymptotic analysis
\begin{equation}\label{MST}
\tau_{\rm on}\approx T_{\rm on} e^{K\Delta S_{\rm on}},
\end{equation}
where \(T_{\rm on}\) is a prefactor which is independent of $K$ in many cases. Although for one dimensional systems the prefactor of MST can be obtained \cite{TPTRate1D}, there are no available results in high dimensions  because  of the geometry problem in more than one spatial dimension and the non-gradient nature of the system \cite{Schuss, Maier}. Therefore we compare the MC simulations with the exponential time part and adjust the prefactor $T_{\rm on}$ to fit the numerical results. Figure \ref{MST:f}(a) and (b) demonstrate the MST versus the variation of parameters \(b\) and \(n_{50}\), respectively. It shows that the MST is excellently predicted by Eq. (\ref{MST}) up to a slowly varying prefactor. The analysis for $\tau_{\rm off}$ is similar. Our results in Fig. \ref{MST:f}(a) show that the MST from both {\it off}-to-{\it on} and {\it on}-to-{\it off} states decrease rapidly when translation rate b is increased.
And when \(n_{50}\) increases in the considered interval,  MST from {\it off} to {\it on} state  increases exponentially, while MST from {\it on} to {\it off} state decreases exponentially. When  $n_{50}$ is about 1004, these two MSTs are equal. These results may provide hints to the range of kinetic reaction rates.

\begin{figure}[tbp]
\includegraphics{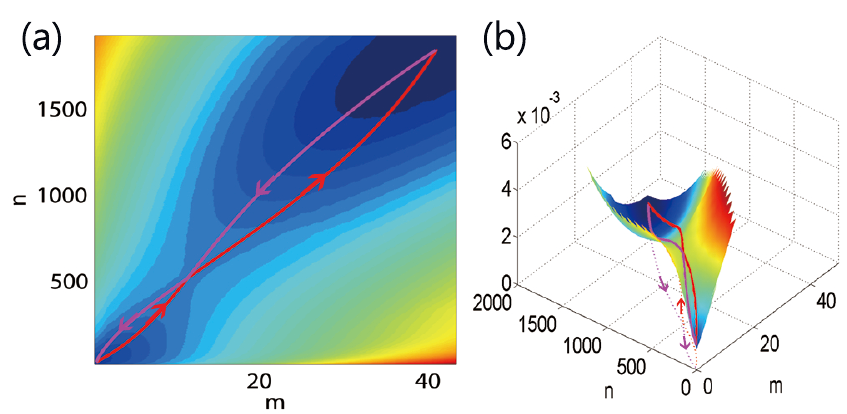}
\caption{\label{potential}(color online). Quasipotential energy landscape of the whole genetic switching system with (a) two and (b) three dimensional view as well as switching paths between two stable fixed points.  Each path passes through the saddle point. Here, \(K=ab=2400, b=22.5, h=2, n_{50}=1000, k_0^{\min}=k_1^{\min}=a/100, k_0^{\max}=k_1^{\max}=a\) and \(\gamma=2.\)}
\end{figure}

Based on the obtained  quasi-potential values for each point from {\it on} and {\it off} states as starting point, we can reconstruct the global quasi-potential energy landscape for genetic switching model. The two and three dimensional view of the quasi-potential energy landscape is shown in Fig. \ref{potential} and the reconstruction details can be found in {\it SM}. The reconstruction for more general stochastic dynamical systems can be referred to \cite{Freidlin1998,Freidlin2000}. In  Fig. \ref{potential}, we observe that the {\it on} and {\it off} states correspond to two local minimum on the quasi-potential energy landscape, the saddle of the deterministic dynamical system exactly corresponds to the saddle point on the quasi-potential energy landscape, too.
%Further investigation shows that both the uphill and downhill paths do not coincide with the steepest ascent or %descent path of the quasi-potential due to the non-gradient nature of the dynamical gentic systems.
As discussed in \cite{Freidlin1998}, we can obtain that the most probable uphill path satisfies $\dot{\boldsymbol \varphi} = \boldsymbol a (\boldsymbol\varphi)\cdot \nabla U(\boldsymbol \varphi) + \boldsymbol l (\boldsymbol \varphi)$, and the most probable downhill path satisfies $\dot{\boldsymbol \psi} = \boldsymbol b (\boldsymbol \psi) = -\boldsymbol a (\boldsymbol\psi)\cdot \nabla U(\boldsymbol \psi) + \boldsymbol l (\boldsymbol \psi)$ if the underlying stochastic dynamics has the form $\dot{\boldsymbol x} = \boldsymbol b (\boldsymbol x) + \sqrt{\varepsilon} \boldsymbol \sigma(\boldsymbol x) \cdot \dot{\boldsymbol w}$ and the drift $\boldsymbol b$ has the decomposition $\boldsymbol b (\boldsymbol x) = -\boldsymbol a(\boldsymbol x)\cdot \nabla U(\boldsymbol x) + \boldsymbol l(\boldsymbol x)$ such that $\boldsymbol l(\boldsymbol x)\cdot \nabla U(\boldsymbol x) = 0$, where $\dot{\boldsymbol w}$ is the standard temporal Gaussian white noise and  the diffusion matrix $\boldsymbol a(\boldsymbol x)=\boldsymbol \sigma( \boldsymbol x)\cdot \boldsymbol \sigma(\boldsymbol x)^{T}$. The mathematical derivations and observations in Fig. \ref{potential} show that the similar picture still holds for the chemical jump processes but the argument by simple orthogonal type decomposition of the drift is no longer valid if we recall that the uphill dynamics is  $\dot{\boldsymbol x} = \nabla_{\boldsymbol p} H(\boldsymbol x, \nabla_{\boldsymbol x}S)$. In general, this form does not permit to specify some matrix $\boldsymbol a$ and make a meaningful decomposition because of nonlinearity. But it is instructive to remark that
$\dot{\boldsymbol x} = \nabla_{\boldsymbol p} H(\boldsymbol x, \nabla_{\boldsymbol x}S) \approx \nabla_{\boldsymbol p} H(\boldsymbol x, 0) + \nabla^{2}_{\boldsymbol {pp}} H(\boldsymbol x, 0) \cdot\nabla_{\boldsymbol x} S$ when $\boldsymbol x$ is close to the critical points with property $\nabla_{\boldsymbol x} S =0$. Based on the fact that $\dot{\boldsymbol x} =  \nabla_{\boldsymbol p} H(\boldsymbol x, 0)$ gives the mean field dynamics, the above derivation actually states that the behavior of the system is well approximated by the chemical Langevin process when the transition path is close to the metastable states and saddle points. More detailed discussions on this point may be referred to {\it SM}.

\begin{figure}[htbp]
\includegraphics{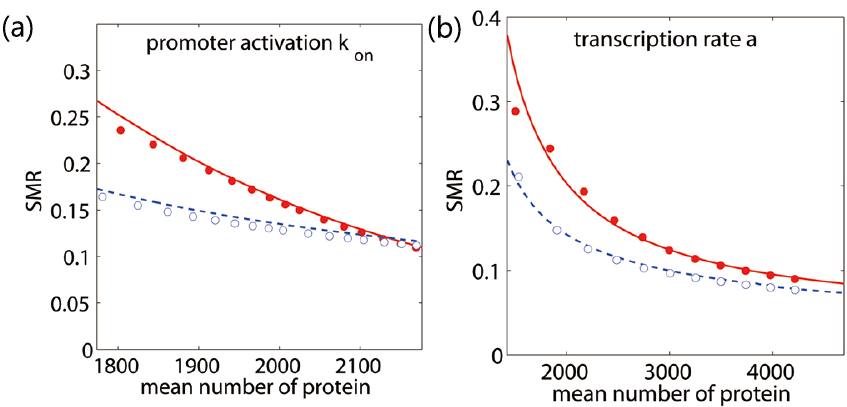}
\caption{\label{SMR_f}(color online). The SMR vs mean number of protein induced by varying promoter activation rate \(k_{on}\) (a) and transcription rate $a$ (b). Analytical results of eukaryotic model (red solid line) and prokaryotic model (blue dashed line), while MC simulations (\(\bullet\)) and (\(\circ\)), respectively. }
\end{figure}

The quasi-potential energy landscape not only provides the pictorial illustration for the dynamical transitions in genetic switching models, it also includes more quantitative information to understand the metastability in gene expressions in single-cell observations \cite{Munsky2012}, such as the fluctuation property like the ratio of the standard deviation to the mean (SMR). The SMR for our genetic switching model can be obtained by the following calculations. From Eqs. \eqref{CMEs} and \eqref{WKB} we know that the contribution of the inactive and active promoter states to the QSD has similar forms $P(x,y),Q(x,y)\sim \exp[-KS(x,y)]$. Thus the whole QSD $\mathcal P(x,y)$ reads $\mathcal P(x,y)=P(x,y)+Q(x,y)\sim e^{-KS(x,y)}$. We can expand $S(x,y)$ in the vicinity of $(x_{on},y_{on})$ up to second order thus get the Gaussian approximation
\begin{equation}\label{Gaussian_distribution t}
  \mathcal P(x,y)\simeq\frac{1}{(2\pi)|\mathbf\Sigma|^{\frac{1}{2}}}\exp\Big\{-\frac12(\mathbf x-\mathbf\mu)\mathbf\Sigma^{-1}(\mathbf x-\mathbf\mu)^T\Big\}.
\end{equation}
Here, $\mathbf x=(x,y), \mathbf\mu=(x_{on},y_{on})$, $\mathbf\Sigma=(S_{ij})_{2\times2}$, and $|\mathbf\Sigma|$ is the determinant of matrix $\mathbf\Sigma$. Eq. \eqref{Gaussian_distribution t} holds only in the vicinity of the {\it on} state with standard deviations $\sigma_m=(KS_{yy}''/|\mathbf\Sigma|)^{\frac12}, \sigma_n=(KS_{xx}''/|\mathbf\Sigma|)^{\frac12}$.
%Here $S_{xx}'', S_{yy}''$ and $S_{xy}''$ are evaluated at $(x_{on},y_{on})$.
With the $\sigma_m$ and $\sigma_n$ above, we can easily obtain the SMR as shown in Fig. \ref{SMR_f}.

 We distinguish among several mechanisms of DNA/promoter activation in genetic switches by selecting different promoter transition rates \(a\) and mRNA degradation rates \(\gamma\), which is similar to the previous work of Raser and O'Shea  \cite{OShea2004}. The above model in Fig. \ref{model} is set with the parameters  \(a\)=106.7 and \(\gamma=2.0\). We set a slow chromatin-remodeling eukaryotic model with \(k_0^{\max}=k_1^{\max}=a/10\), \(k_0^{\min}=k_1^{\min}=a/1000\), and \(\gamma=1.5\), while a prokaryotic model is set as \(k_0^{\max}=k_1^{\max}=10a\), \(k_0^{\min}=k_1^{\min}=a/10\), with a higher mRNA degradation rate \(\gamma=15\). The results in Fig. \ref{SMR_f} show that the noise level in eukaryotic model is almost always bigger than that in prokaryotic model. And the relationship between SMR and mean expression level is different from the no-feedback models \cite{OShea2004}, especially for the changing of transcription rate \(a\). It is interesting to find that when increasing the transcription rate \(a\), noise level of {\it on} state decreases in the model with positive feedback, while protein noise increases in one-state chromatin-remodeling eukaryotic model without positive feedback (Case I in Raser and O'Shea's work). In {\it SM}, we also compare the SMR curves of different promoter transition rates  \(a\) with fixed \(\gamma=2.0\) (Fig. S4) and provide the stochastic features of mRNA (Fig. S3).
\begin{figure}[htbp]
\includegraphics{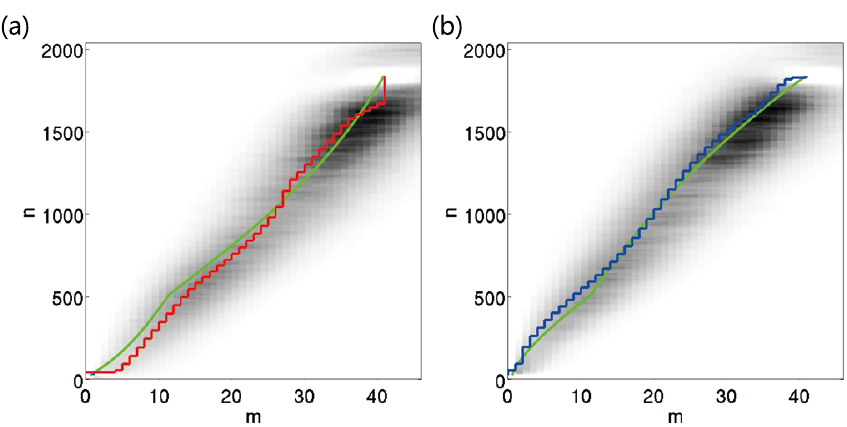}
\caption{\label{fig:TPT}(color online). Switching paths compared with gMAM result. gMAM result is colored green in both subfigure. (a) from {\it off} to {\it on} state and (b) from {\it on} to {\it off} state and MC simulations for both switching trajectories. We take the nearest 4 lattice points around starting and ending point in gMAM as the starting and ending set. Red line means the system is in DNA active state while blue line means DNA is inactive. The parameter setup is the same as in  Fig. \ref{model}.}
\end{figure}

Our methodology above is based on large deviation theory, which requires that the system size tends to infinity. For more general and realistic cases, the system size is finite, thus the transition path theory (TPT) can be applied to find the most probable transition paths \cite{TPT}. Here we choose the nearest 4 lattice points around starting and ending point in gMAM as our starting and ending sets and perform the TPT computations (see Fig. \ref{fig:TPT}). The implementation details can be found in {\it SM}. Fig. \ref{fig:TPT} shows that TPT result fits MC simulation slightly better than gMAM. TPT result in Fig. \ref{fig:TPT}(a) also provides interesting details to understand the uphill trajectory from {\it off} to {\it on} state. In the beginning the {\it off}-to-{\it on} switch, the system is mostly at the active DNA/promoter state thus the mRNA level increases before the translation of protein. While in the beginning of {\it on}-to-{\it off} switch in Fig. \ref{fig:TPT}(b), we observe opposite mechanism with inactive DNA/promoter and first decrease of mRNA. This explains why the uphill transition path from {\it off} to {\it on} state is convex and it is concave from {\it on}-to-{\it off} state.

In this letter, we have presented an analytical framework to reconstruct the quasi-potential energy landscape of genetic switching system while explicitly taking mRNA noise into account. We further analyze the properties of the transition paths and clarify the relation with the gradient-curl decomposition in the previous literatures. This global potential, which is a rationalized version of Waddington potential,  provides a quantitative tool to understand the metastability in more general biological processes with intrinsic noise. The applications to other biological systems such as complex cellular decision making process and the development process of cells will be investigated in the future.

The authors are grateful to Weinan E, Xiang Zhou, Qi Ouyang and Hongli Wang for helpful discussions. The work is supported by NSFC grants no.11174011, 11021463 (F.Li), 11171009 and 91130005 (T.Li).

\bibliographystyle{unsrt}

\end{document}